\documentclass[sigconf,natbib=true,anonymous=false]{acmart}
\AtBeginDocument{%
  }

\copyrightyear{2025}
\acmYear{2025}
\setcopyright{cc}
\setcctype{by}
\acmConference[SIGIR '25]{Proceedings of the 48th International ACM SIGIR Conference on Research and Development in Information Retrieval}{July 13--18, 2025}{Padua, Italy}
\acmBooktitle{Proceedings of the 48th International ACM SIGIR Conference on Research and Development in Information Retrieval (SIGIR '25), July 13--18, 2025, Padua, Italy}\acmDOI{10.1145/3726302.3730162}
\acmISBN{979-8-4007-1592-1/2025/07}

\definecolor{midnightgreen}{rgb}{0.0, 0.29, 0.33}

\usepackage{tabularx}

\begin{document}


% THIS IS A SUGGESTION to better (and clearly) reflect the idea of the paper.
\title{Aligning Web Query Generation with Ranking Objectives via \\Direct Preference Optimization}
% The previous title is still here

%\title{Guiding Query Generation Towards Better Web Retriever Training}

% ----
%\title{Retriever Training Oriented Synthetic Query Generation}

% \title{Improving Synthetic Queries for Training Web Retrieval Models with Direct Preference Optimization}

\author{João Coelho}
\affiliation{%
  \institution{Carnegie Mellon University}
  \city{Pittsburgh}
  \state{Pennsylvania}
  \country{United States}
  }
% \affiliation{%
%   \institution{Instituto Superior Técnico and INESC-ID, University of Lisbon, Portugal}}
\email{jmcoelho@andrew.cmu.edu}

\author{Bruno Martins}
\affiliation{%
  \institution{Instituto Superior Técnico and INESC-ID}
    \city{Lisbon}
    \country{Portugal}
  }
  \email{bruno.g.martins@tecnico.ulisboa.pt}

\author{João Magalhães}
\affiliation{%
  \institution{NOVA School of Science and Technology}
   \city{Lisbon}
   \country{Portugal}
    }
\email{jmag@fct.unl.pt}

\author{Chenyan Xiong}
\affiliation{%
  \institution{Carnegie Mellon University}
  \city{Pittsburgh}
  \state{Pennsylvania}
  \country{United States}
  }
\email{cx@cs.cmu.edu}

\renewcommand{\shortauthors}{João Coelho, Bruno Martins, João Magalhães, and Chenyan Xiong}

\begin{abstract}
Neural retrieval models excel in Web search, but their training requires substantial amounts of labeled query-document pairs, which are costly to obtain. With the widespread availability of Web document collections like ClueWeb22, synthetic queries generated by large language models offer a scalable alternative. Still, synthetic training queries often vary in quality, which leads to suboptimal downstream retrieval performance. Existing methods typically filter out noisy query-document pairs based on signals from an external re-ranker. In contrast, we propose a framework that leverages Direct Preference Optimization (DPO) to integrate ranking signals into the query generation process, aiming to directly optimize the model towards generating high-quality queries that maximize downstream retrieval effectiveness. 
Experiments show higher ranker-assessed relevance between query-document pairs after DPO, leading to stronger downstream performance on the MS~MARCO benchmark when compared to baseline models trained with synthetic data.
\end{abstract}





\maketitle

\section{Introduction}

The effective training of neural retrieval models benefits from large-scale query-document relevance annotations~\citep{DBLP:conf/sigir/FangZAMS0024}. However, acquiring such data remains a bottleneck due to its high costs. To address this limitation, recent work has explored using Large Language Models (LLMs) to generate synthetic queries from documents~\citep{DBLP:conf/iclr/DaiZMLNLBGHC23,DBLP:journals/corr/abs-2403-20327,DBLP:conf/acl/WangYHYMW24}, enabling retrieval models to be trained without full reliance on human annotations. Combined with the availability of large Web corpora such as ClueWeb22~\citep{DBLP:conf/sigir/OverwijkXC22}, this establishes LLM query generators as a scalable approach for training data creation. Despite this potential, the quality of synthetic queries varies significantly, with studies showing that it still lags behind human-annotated data~\citep{DBLP:conf/sigir/FangZAMS0024}, which degrades downstream effectiveness~\citep{DBLP:conf/ecir/GospodinovMM23}. 

To mitigate this problem, approaches for query generation consider post hoc filtering, e.g., using an external ranking signal~\citep{DBLP:journals/corr/abs-2301-01820} or through consistency checking during negative sampling~\citep{DBLP:conf/iclr/DaiZMLNLBGHC23}. While this improves query quality, it does not guide the query generator toward producing queries that maximize retrieval performance.

In this work, instead of post hoc query filtering, we propose a framework that directly integrates ranking feedback into the query generation process using Direct Preference Optimization (DPO)~\citep{DBLP:conf/nips/RafailovSMMEF23}. To this end, we generate multiple queries per document, score them using a ranking model, and build a preference dataset to optimize the query generator toward producing high-quality queries. This approach enables the model to learn directly from ranking signals, aiming to improve the overall quality of queries used for dense retrieval training. More specifically, we independently consider ranking rewards from point-wise re-rankers and list-wise LLM prompting techniques, demonstrating the framework's flexibility and opening future directions for reward selection.

We conduct experiments on the MS MARCO dataset~\citep{DBLP:conf/nips/NguyenRSGTMD16}, i.e. a standard benchmark for passage and document retrieval, to evaluate the impact of DPO-optimized queries on retrieval performance. Our framework demonstrates three key contributions: (1) a training paradigm that directly aligns query generation with ranking objectives, diminishing the need for post hoc filtering inefficiencies due to increased ranker-assessed relevance; (2) flexibility in integrating diverse ranking signals, including both point-wise re-rankers and LLM-based list-wise prompting; and (3) empirical validation, showing that retrievers trained with synthetic data from our generation pipeline outperform other open-source alternatives in the Web domain. Our models and code are available in a public GitHub repository: \href{https://github.com/cxcscmu/dpo-query-generation}{https://github.com/cxcscmu/dpo-query-generation}.
\section{Related Work}

Transformer-based bi-encoders are the standard architecture for dense retrieval~\citep{DBLP:conf/iclr/XiongXLTLBAO21,DBLP:conf/emnlp/KarpukhinOMLWEC20}, typically trained with contrastive objectives and hard negative mining strategies such as ANCE~\citep{DBLP:conf/iclr/XiongXLTLBAO21}. Retrieval-aligned pre-training on in-domain corpora is also commonly adopted to improve retrieval effectiveness~\citep{DBLP:conf/acl/GaoC22, lu-etal-2021-less, DBLP:conf/emnlp/XiaoLSC22, DBLP:conf/acl/LeeCT19, DBLP:conf/sigir/MaGZFC22, DBLP:journals/corr/abs-2401-11248}.

% Unsupervised dense retrieval methods still leverage contrastive training methodologies, but without relying on human-annotated labels. Multiple approaches have been explored to obtain positive samples, for instance, based on anchor text~\citep{DBLP:conf/sigir/XieLX23}, sampling positive spans from a single document~\citep{DBLP:conf/acl/LeeCT19, DBLP:journals/tmlr/IzacardCHRBJG22}, or heuristically mining pairs from structured documents (e.g., title-paragraph)~\citep{DBLP:journals/corr/abs-2212-03533}.

The community has explored generating synthetic queries from documents for model training. Approaches such as Doc2Query~\citep{DBLP:journals/corr/abs-1904-08375} and DocT5Query~\citep{nogueira2019doc2query} train a lightweight Transformer~\citep{DBLP:conf/nips/VaswaniSPUJGKP17} on labeled query–document pairs to expand document representations. Subsequent work~\citep{DBLP:conf/ecir/GospodinovMM23} demonstrates that filtering out hallucinated queries can further enhance downstream performance. More recent methods leverage LLMs. For instance, InPars~\citep{DBLP:journals/corr/abs-2202-05144} employs few-shot prompting with filtering based on generation probability, while InPars-v2~\citep{DBLP:journals/corr/abs-2301-01820} leverages a supervised ranker for query filtering. Similarly, Promptagator~\citep{DBLP:conf/iclr/DaiZMLNLBGHC23} uses task-specific prompts to perform few-shot query generation. Building on these approaches, the Gecko model~\citep{DBLP:journals/corr/abs-2403-20327} iteratively refines synthetic queries through a process involving retrieval, re-ranking, positive relabeling, and hard negative sampling, thereby improving the quality of the training data.

%\cx{not very relevant?}
% Other lines of work explore synthesizing documents instead of queries. Methods such as Syntriever~\citep{kim2025syntriever} generate synthetic documents (both positive and negative) conditioned on input queries, framing this as knowledge distillation from LLMs. Other approaches generate fully synthetic query-document pairs at scale, enabling retrieval-oriented pre-training~\citep{DBLP:conf/acl/WangYHYMW24}.

Recent work has explored reinforcement learning to improve synthetic query generation. Token-level Proximal Policy Optimization (TPPO)~\citep{DBLP:journals/corr/abs-2411-00722, DBLP:journals/corr/SchulmanWDRK17} has been applied to query suggestion tasks, optimizing generation based on token-level rewards derived from user interaction histories. While this setting targets interactive scenarios rather than offline generation from documents, it highlights the promise of reinforcement learning for enhancing query quality.

% todo: add token-level PPO paper

\section{Methodology}

This section describes our query generation process, and details the training of retrieval models using the resulting synthetic queries.

\subsection{Query Generation}
\label{sec:qgen}
%In the Web retrieval domain, the task of synthetic query generation aims to produce plausible search queries, ideally similar to those that users might issue to retrieve a given document. 
We start from a baseline query generator $\mathcal{M}_g$ (e.g., an LLM of choice or an existing model fine-tuned for query generation). To synthesize queries, we follow a contrastive prompting strategy~\citep{DBLP:journals/corr/abs-2011-01580} where given a document $d$, a negative example $d^-$ is sampled from a corpus. Then, the model is prompted to generate a query $q$ that is specific to $d$, while being less relevant to $d^-$, i.e. $q \sim \mathcal{M}_g(d, d^-)$. For simplicity, we represent this generation process as $\mathcal{M}_g(d)$, assuming implicit contrastive prompting. 

By aiming to increase $(q,d)$ similarity, this strategy mitigates retrieval training scenarios where poorly generated queries allow other corpus documents to dominate over the original positive during negative sampling, which is a source of noise. 

\subsection{Aligning the Query Generator}
\label{sec:align}
Instead of filtering queries generated by $\mathcal{M}_g$, we hypothesize that the query generator can be guided towards better downstream retrieval performance.  To this end, we propose leveraging a reward model to obtain a preference dataset for DPO alignment. Given the baseline generator $\mathcal{M}_g$ and a corpus $\mathcal{C}$, we start by sampling a random pool of documents from $\mathcal{C}$, denoted as $\mathcal{D}$. For each document in $\mathcal{D}$, we use $\mathcal{M}_g$ to generate a set of $n$ synthetic queries: 
\begin{equation}
   Q_d = \{q_{d,1}, \dots, q_{d,n} \} \sim \mathcal{M}_g(d),\;\forall d \in \mathcal{D}. 
\end{equation}

\noindent\textbf{Obtaining Rewards:} As a first reward mechanism, we consider the usage of a ranking model $\mathcal{R}$ to score each individual pair, since previous work has established this as a strong filtering method~\citep{DBLP:journals/corr/abs-2301-01820}. 
More specifically, we used \texttt{bge-reranker-v2-m3}\footnote{\url{https://huggingface.co/BAAI/bge-reranker-v2-m3}}, i.e. a small cross-encoder model available on HuggingFace, which offers a good trade-off between retrieval performance and computation resources. 

By applying $\mathcal{R}$ on the synthetic pairs, we obtain triplets in the form 
$\{(q, d, \mathcal{R}(q, d)) \mid d \in \mathcal{D}, q \in Q_d \}$. We can then extract a preference dataset $\mathcal{D}_\text{rr}$, where, for each document, a random pair of queries is sampled such that one has a higher score than the other:
\begin{equation}
\mathcal{D}_\text{rr} = \{(d, q^+, q^-) \mid d \in \mathcal{D},\; (q^+, q^-) \in Q_d,\; \mathcal{R}(q^+, d) > \mathcal{R}(q^-, d) \}.
\end{equation}

Alternatively, we can consider list-wise reward models, for instance through GPT3.5 prompting, which previous work has established as a competitive ranking method~\citep{DBLP:conf/emnlp/0001YMWRCYR23}. In this setup, given $d$ and $Q_d$, the model $\mathcal{G}$ is prompted to explicitly select the best and worst queries from $Q_d$, producing a preference dataset $\mathcal{D}_{gpt}$:
\begin{equation}
\mathcal{D}_\text{gpt} = \{(d, q^+, q^-) \mid d \in \mathcal{D},\; (q^+, q^-) \sim \mathcal{G}(d, Q_d) \}.
\end{equation}

\noindent\textbf{Fine-tuning the Generator:} Both $\mathcal{D}_\text{rr}$ and $\mathcal{D}_\text{gpt}$ allow training $\mathcal{M}_g$ following the Direct Preference Optimization (DPO) loss~\citep{DBLP:conf/nips/RafailovSMMEF23}: 
{\scriptsize
\begin{equation}
\label{eq:dpo-loss}
\mathcal{L} = -\mathbb{E}_{(d,q^+,q^-)\sim \mathcal{D}_\text{pref} }\left[\log\sigma\left(\beta\log\frac{\mathcal{M}_g^*(q^+ \mid d)}{\mathcal{M}_g(q^+ \mid d)}-\beta\log\frac{\mathcal{M}_g^*(q^- \mid d)}{\mathcal{M}_g(q^- \mid d)}\right)\right] \;,
\end{equation}
}

\noindent where $\mathcal{D}_\text{pref}$ is the preference dataset (e.g., $\mathcal{D}_\text{rr}$ or $\mathcal{D}_\text{gpt}$), and $\beta$ is an hyper-parameter for tuning. This aims to ensure that the likelihood of generating ranker-preferred queries is higher, thereby directly including the ranking signals into the generation process rather than as a filtering stage, which we argue contributes to more robust large-scale generation.

\subsection{Dense Retriever Training}
\label{dr_training}
In order to obtain positive pairs for dense retriever training, we can use the aligned query generator $\mathcal{M^*}_g$, to conduct inference on a target Web document corpus. Given a random sample of $k$ documents, $\mathcal{D}_k$, we use $\mathcal{M^*}_g$ to sample one query per document, obtaining a set of synthetic positive pairs:
\begin{equation}
    \mathcal{D}_\text{train} = \{(q, d) \mid d \in \mathcal{D}_k,\; q \sim \mathcal{M^*}_g(d)  \} \;.
\end{equation} 

For the embedding model ($\mathcal{M}_e$), we leverage Qwen2.5-0.5B~\citep{DBLP:journals/corr/abs-2412-15115} as the language model backbone. Following previous work on decoder-only dense retrievers, we apply a bi-directional attention mask and conduct masked next token prediction as continual pre-training. This has been shown to better align the model to the task of generating dense representations, with continual pre-training being necessary for the model to adapt to the new attention mask~\citep{DBLP:journals/corr/abs-2404-05961}. We leverage 100k random documents from MS~MARCO~\citep{DBLP:conf/nips/NguyenRSGTMD16} for pre-training, as a form of domain adaptation~\citep{DBLP:conf/acl/GaoC22}.

\begin{table*}[t!]
  \centering
  \caption{Performance of multiple models trained with synthetic data on Web benchmarks, including model size and number of training pairs. $^*$Synthetic-only checkpoint unavailable for reproduction; MARCO passage result extracted from original paper.}
  \label{tab:performance-config}
%  \resizebox{\textwidth}{!}{
    \begin{tabular}{l @{\hskip 3em}  c c @{\hskip 3em} cccc}
      \toprule
       &  &  & \multicolumn{4}{c}{Datasets} \\
      \cmidrule(lr){4-7}
                   &   &       & MARCO Doc & MARCO Passage & TREC-DL 19 Passage & TREC-DL 20 Passage \\
       Model                 & Size            & Pairs          & \scriptsize{MRR@100}  & \scriptsize{nDCG@10} & \scriptsize{nDCG@10}   & \scriptsize{nDCG@10}   \\
      \midrule
      BM25~\citep{DBLP:journals/ftir/RobertsonZ09}       & -  & -  & 0.2300      & 0.2280   & 0.5058      & 0.4796 \\
      Contriever~\citep{DBLP:journals/tmlr/IzacardCHRBJG22} & 0.1B  & 700M  & 0.2322      & 0.2060   & 0.4794      & 0.4352 \\
      WebDRO~\citep{han2024enhancingdenseretrieversrobustness}      & 0.2B  & 13.8M  & 0.3030      & 0.2741   & 0.4981      & 0.4591 \\
      $\text{E5}_{\text{large}}$\citep{DBLP:journals/corr/abs-2212-03533} & 0.3B  & 270M & 0.3039      & 0.2620   & 0.5108      & 0.5024 \\
      MiniCPM-Embedding\footnotemark[2] & 2.4B  & 2M  & 0.3180     & 0.2980   & 0.5730      & \textbf{0.5646} \\
      $\text{E5}_{\text{Mistral}}$\citep{DBLP:conf/acl/WangYHYMW24}$^*$ & 7B  & 0.5M  & -      & 0.2570   & -      & - \\
      \midrule 
      Ranker-DPO         & 0.5B  & 1M  & \underline{0.3727} & \underline{0.3286}   & \underline{0.6171} & 0.5337 \\
      GPT-DPO            & 0.5B  & 1M  & \textbf{0.3795} & \textbf{0.3340}   & \textbf{0.6213} & \underline{0.5496} \\
      \bottomrule
    \end{tabular}
%  }
\end{table*}

Given $\mathcal{D}_\text{train}$, we follow a standard setup, where an external model is used to sample hard negatives for each query. More specifically, given the top-100 documents for a query, 5 negatives are sampled from the documents that are ranked below the original positive~\citep{DBLP:journals/corr/abs-2405-17428}. We also consider the relabeling mechanism proposed by Gecko~\citep{DBLP:journals/corr/abs-2403-20327}, where if the original positive document is not in the top-100, the top-1 is considered as the new positive document instead of discarding the query.

We fine-tune $\mathcal{M}_e$ leveraging in-batch and cross-GPU negatives, aiming to minimize the InfoNCE loss~\citep{DBLP:journals/corr/abs-1807-03748}:
%{\scriptsize
\begin{equation}
\label{eq:contrastive-loss}
\mathcal{L} = - \log \frac{e^{\mathrm{cos}(\mathcal{M}_e(q), \mathcal{M}_e(d))}}{e^{\mathrm{cos}(\mathcal{M}_e(q), \mathcal{M}_e(d))} + \sum_{n \in \mathcal{N}} e^{\mathrm{cos}(\mathcal{M}_e(q), \mathcal{M}_e(n))}} \;,
\end{equation}
%}

\noindent where $\mathcal{N}$ is the set of all negatives. Representations $\mathcal{M}_e(.)$ are obtained by average pooling and compared using the cosine similarity.

\section{Experiments}

This section describes our experimental setup, evaluation, and results. We begin by detailing the configurations for both query generation and retrieval training, along with an overview of the evaluation benchmarks. Next, we present the experimental results, assessing the impact of our query generation strategy and conducting scaling experiments in terms of the number of documents.

\subsection{Setup and Implementation Details}

For our corpus $\mathcal{C}$, we consider ClueWeb22-B~\citep{DBLP:conf/sigir/OverwijkXC22}, which contains the most popular and frequently visited web pages of ClueWeb22, making it well-suited for query generation due to its inherent informativeness. We leverage a sample of 35 million documents to ensure broad coverage while maintaining computational efficiency.

For query generation, we leverage an open-source model as our baseline $\mathcal{M}_g$, which was used to obtain the synthetic data to train the MiniCPM-Embedding model\footnote{\label{note1}\url{https://huggingface.co/openbmb/MiniCPM-Embedding}}. This generator was obtained by fine-tuning LLaMA-3-7B~\citep{DBLP:journals/corr/abs-2407-21783} with synthetic query-document pairs obtained from GPT-4, following the contrastive prompting strategy described in Section~\ref{sec:qgen}.

Starting with a pool of 100k random documents from $\mathcal{C}$, we sample 5 queries per document using $\mathcal{M}_g$, and use this data to follow the process described in Section~\ref{sec:align}, in order to obtain the improved generator $\mathcal{M}^*_g$. The document count (100k) was selected as a practical upper bound that allows model convergence within reasonable training cost. Due to the size of the model, DPO training is conducted using LoRA~\citep{DBLP:conf/iclr/HuSWALWWC22}, considering 8 NVIDIA L40S GPUs, a batch size of 8, and a learning rate of 1e-5. This setup induces a one-time 10-hour overhead compared to just using $\mathcal{M}_g$. 

After obtaining the aligned query generator, we randomly sample documents from $\mathcal{C}$ (excluding the previous 100k documents used for model training) and generate one query per document. The obtained query-document positive pairs are then used to train two retrieval models following the setup introduced in Section~\ref{dr_training}, independently leveraging synthetic data obtained from aligning the generator with ranker or GPT rewards. We adopt the same embedding-based negative sampling strategy used in MiniCPM-Embedding, by using an external model to obtain negatives from $\mathcal{C}$. Training is conducted for one epoch with a batch size of 64, a learning rate of 1e-4, and distributed across 8 L40S NVIDIA GPUs.

\subsection{Evaluation}

For evaluation, we leverage MS~MARCO~\citep{DBLP:conf/nips/NguyenRSGTMD16}, i.e. a standard Web benchmark including both document and passage retrieval, as well as the queries from the TREC-DL 2019 and 2020 passage tasks. Regarding metrics, we follow common practice: for MS~MARCO document retrieval we use the Mean Reciprocal Rank (MRR)~\citep{DBLP:reference/db/Craswell09a}; for MS-MARCO passage retrieval and associated TREC-DL tracks~\citep{DBLP:journals/corr/abs-2003-07820, DBLP:conf/trec/CraswellMMYC20}, we use the normalized Discounted Cumulative Gain (nDCG)~\citep{DBLP:journals/tois/JarvelinK02}. 

\subsection{Retrieval Performance}

Table~\ref{tab:performance-config} presents a comparative evaluation of our models against open-source approaches leveraging synthetic data. Specifically, we benchmark our models against several baselines: lexical retrieval (BM25), models trained on anchor data (WebDRO), models trained on heuristically selected pairs from structured documents (E5\textsubscript{large}), and models trained on LLM-generated data (E5\textsubscript{Mistral} and MiniCPM-Embedding). For MiniCPM-Embedding, we report results obtained using synthetic data exclusively, noting that this synthetic data was generated by the same generator employed as a baseline in this study. Overall, our methods  (i.e., either using a ranker or GPT as reward models) yield superior performance on Web benchmarks, showcasing the effectiveness of the proposed framework. Both reward mechanisms demonstrate similar performance, likely due to the fact that both rely on semantic cues when distinguishing between queries. The next sections analyze query quality and document scale to elucidate the observed improvements.

\begin{table}[t!]
\centering
\caption{Comparison between the baseline generator and the DPO variant, considering a pool of 250K generated queries and consistency filtering.}
\label{tab:comparison}
\begin{tabular}{@{}lcccc@{}}
\toprule
               &  &               & \multicolumn{2}{c}{MARCO Doc}                            \\
\cmidrule(lr){4-5}
               & Initial Pool             &   Kept                & MRR@100   & R@1000    \\ \midrule
Before DPO     & 250K         & 156K (62\%)       & 0.1635    & 0.5902    \\
After DPO      & 250K         & 231K (92\%)       & 0.2625    & 0.7491    \\ \bottomrule
\end{tabular}
\end{table}

% \begin{table}[t]
% \centering
% \caption{Comparison of queries before and after DPO.}
% \label{tab:queries}
% \footnotesize
% \setlength{\tabcolsep}{5pt}
% \begin{tabularx}
% {\columnwidth}{@{}>{\centering\arraybackslash}X >{\centering\arraybackslash}X@{}}

% \toprule
% \textbf{Before DPO} & \textbf{After DPO} \\
% \midrule
% What are the steps to adjust the idle on a 2003 Toyota Tacoma 2.4 liter? 
%   & Adjusting idle on 2003 Toyota Tacoma \\
% What makes the 2019 Ram 1500's frame and hybrid system superior?
%   & 2019 Ram 1500 redesign and features \\
%   How do religious beliefs influence faith and practice during pandemic crises like COVID-19? & Impact of COVID-19 on religion and faith perception \\
% \bottomrule
% \end{tabularx}
% \end{table}

\renewcommand{\arraystretch}{1.5} 

\begin{table}[t]
\centering
\small
\caption{Comparison of queries generated before and after DPO. Each pair was generated for the same document.}
\label{tab:queries}
\small
\setlength{\tabcolsep}{5pt}
\begin{tabularx}{\columnwidth}{@{}>{\centering\arraybackslash}X >{\centering\arraybackslash}X@{}}
\toprule
\textbf{Before DPO} & \textbf{After DPO} \\
\midrule
What are the steps to adjust the idle on a 2003 Toyota Tacoma 2.4 liter? 
  & Adjusting idle on 2003 Toyota Tacoma \\
What makes the 2019 Ram 1500's frame and hybrid system superior?
  & 2019 Ram 1500 redesign and features \\
How do religious beliefs influence faith and practice during pandemic crises like COVID-19?
  & Impact of COVID-19 on religion and faith perception \\
\bottomrule
\end{tabularx}
\end{table}

\subsection{Query Retention}

To evaluate the impact of generator alignment via DPO, we analyze synthetic query quality through query retention rate and ranker-assessed relevance. We focus on the ranker-DPO variant, as its alignment reward model $\mathcal{R}$ provides relevance scores that enable comparisons between baseline and optimized queries. We generate 250k queries, before and after DPO, and train two dense retrieval models. Negative documents are sampled from the previously introduced corpus of 35 million candidates, making the generation task more demanding: suboptimal queries often fail to distinguish the original positive from high-relevance negatives introduced by large-scale sampling, which can contribute to training noise.

%This large-scale negative sampling presents a significant challenge to query generation efficacy: suboptimal generators may produce queries with insufficient discriminative power to distinguish the original positive document from the negative pool, which, given its size, increases the probability of higher-relevance matches existing among the sampled negatives, introducing noise to the training process. 

Table~\ref{tab:comparison} shows that after DPO, following a standard consistency parsing mechanism that discards queries if the original positive document is not in the top-100 when sampling negatives, queries have an approximately 50\% higher retention rate. This high difference indicates that integrating ranking feedback via DPO guides the query generation process toward higher relevance, hence improving both data efficiency and quality, given the increase in retrieval performance and query retention.

To further consolidate these findings, we examined the distribution of ranker reward scores $\mathcal{R}(q, d)$ for a sample of 10k query-document pairs. As depicted in the KDE plot in Figure~\ref{fig:kde}, the post-DPO score distribution exhibits a pronounced rightward shift relative to the one from the base generator. This reflects an improvement in query–document relevance, thereby validating the efficacy of our DPO framework in steering the generator toward higher-quality outputs. As illustrated in Table~\ref{tab:queries}, the DPO-aligned generator produces more concise and web-like queries, transforming verbose questions into focused search terms while preserving the core information intent.

\subsection{Scaling Synthetic Data}

\begin{figure}[t]
    \centering
    \begin{minipage}{0.47\columnwidth}
        \centering
        \includegraphics[width=\textwidth,height=3.7cm,keepaspectratio]{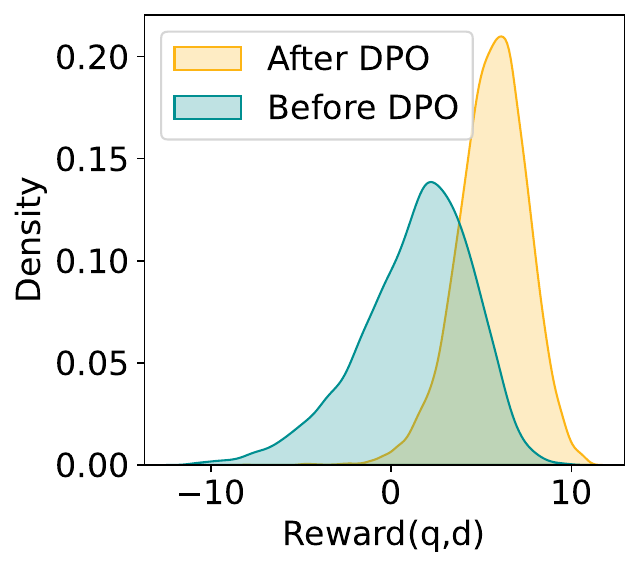}
        \caption{Reward ($\mathcal{R}$) distributions for 10k $(q,d)$ pairs,  before and after DPO.}
        \label{fig:kde}
    \end{minipage}
    \hfill
    \begin{minipage}{0.47\columnwidth}
        \centering
        \includegraphics[width=\textwidth,height=3.7cm,keepaspectratio]{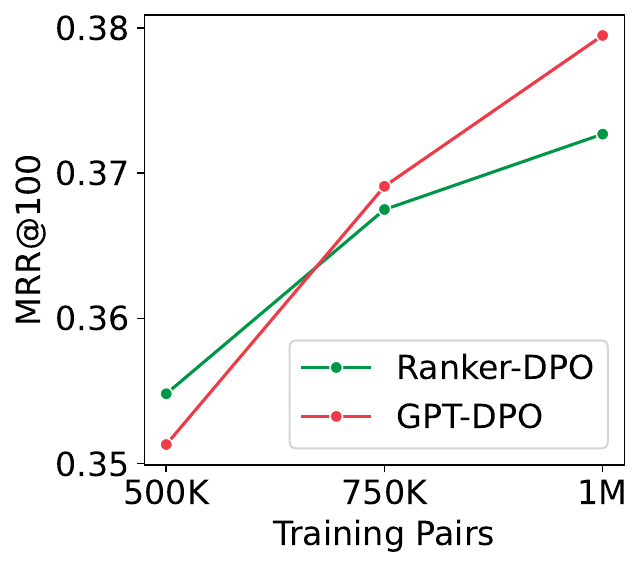}
        \caption{Performance on MARCO Doc as synthetic training pairs increase.}
        \label{fig:lchart}
    \end{minipage}
    %\vspace{-0.3cm}
\end{figure}

We also present an empirical investigation of how model performance scales with synthetic training pairs. Figure~\ref{fig:lchart} demonstrates that both models exhibit progressive improvements in retrieval effectiveness as the document pools expand. While larger document pools consistently enhance model effectiveness, our results show that resource-constrained implementations remain viable, as competitive performance levels can be achieved under limited computational budgets, while maintaining the flexibility to scale.

% \begin{figure}[t]
%     \centering
%     \includegraphics[width=\columnwidth,height=3.9cm,keepaspectratio]{images/kde_plot_square.pdf}
%     \caption{Reward ($\mathcal{R}$) score distributions for 10000 query-document pairs, obtained before and after DPO.}
%     \label{fig:kde}
% \end{figure}

% \begin{figure}[t]
%     \centering
%     \includegraphics[width=\columnwidth,height=3.9cm,keepaspectratio]{images/line_chart_square.pdf}
%     \caption{Performance on MARCO (D) of models trained with varying numbers of synthetic training pairs.}
%     \label{fig:lchart}
% \end{figure}

\section{Conclusions and Future Work}

% \cx{use conclusion to add something new, like what is the high level view from you on the role of this work, and potential impact. rather than repeating abstract}

This paper introduces a framework that aimed at improving synthetic training data generation for dense retrieval, representing a shift from post hoc filtering methods. 
While our implementation focuses on Web retrieval, the principle of using preference optimization to align data generation with task performance can be applied to other systems that rely on synthetic data. Rather than treating query generation and filtering as separate steps, our results suggest that explicitly integrating ranking objectives into the generator training process leads to more robust and data-efficient systems. The empirical improvements we observe validate the proposed approach, pointing toward a promising direction for developing more effective synthetic data generators for dense retrieval training.

% By integrating Direct Preference Optimization (DPO) into the query generation process, we demonstrate how ranking signals can directly shape the generation of synthetic data, leading to more effective retrieval models, and higher data efficiency, by diminishing the need for post-hoc filtering. Our empirical analysis validates our method, showing higher query retention rates and stronger ranked-assessed similarity after DPO, which lead to better performance on the MS~MARCO benchmark.

% This paper addresses the challenge of improving synthetic queries for training dense retrieval models by integrating Direct Preference Optimization (DPO) into the query generation process. Traditional methods rely on post hoc filtering of generated queries. By contrast, the proposed framework leverages ranking feedback to construct preference datasets, enabling the query generator to prioritize queries that maximize downstream retrieval effectiveness. Experimental results on MS~MARCO benchmarks demonstrate improvements, with DPO-optimized queries yielding superior retrieval performance compared to baseline methods. 

Future work may focus on rewards beyond the use of ranker and LLM-based ranking signals, e.g. by incorporating task-specific signals to promote generalization. In addition, while we currently employ a single model for query generation, using multiple models could enhance the diversity of synthetic data~\citep{DBLP:journals/corr/abs-2407-21783} and improve retrieval fine-tuning. Finally, considering the computational cost of query generation, developing smaller and more efficient generators without compromising quality is a worthwhile future direction.

% Extending our approach to non-English retrieval tasks is another appealing direction, given the rising demand for high-quality multilingual systems~\citep{DBLP:journals/corr/abs-2404-08071}.

\begin{acks}
We thank the anonymous reviewers for their valuable comments and suggestions. This research was supported by the Portuguese Recovery and Resilience Plan through project C645008882-00000055 (i.e., the Center For Responsible AI), and also by the Fundação para a Ciência e Tecnologia, specifically through the project with reference UIDB/50021/2020, the project with reference UIDP/04516/2020, and through the Ph.D. scholarship with reference PRT/BD/153683/2021 under the Carnegie Mellon Portugal Program. 
\end{acks}

\bibliographystyle{ACM-Reference-Format}
\bibliography{main}


\begin{thebibliography}{39}

%%% the Web address.

\ifx \showCODEN    \undefined \def \showCODEN     #1{\unskip}     \fi
\ifx \showISBNx    \undefined \def \showISBNx     #1{\unskip}     \fi
\ifx \showISBNxiii \undefined \def \showISBNxiii  #1{\unskip}     \fi
\ifx \showISSN     \undefined \def \showISSN      #1{\unskip}     \fi
\ifx \showLCCN     \undefined \def \showLCCN      #1{\unskip}     \fi
\ifx \shownote     \undefined \def \shownote      #1{#1}          \fi
\ifx \showarticletitle \undefined \def \showarticletitle #1{#1}   \fi
\ifx \showURL      \undefined \def \showURL       {\relax}        \fi
% invisible to TeX
\providecommand\bibfield[2]{#2}
\providecommand\bibinfo[2]{#2}
\providecommand\natexlab[1]{#1}
\providecommand\showeprint[2][]{arXiv:#2}

\bibitem[BehnamGhader et~al\mbox{.}(2024)]%
        {DBLP:journals/corr/abs-2404-05961}
\bibfield{author}{\bibinfo{person}{Parishad BehnamGhader}, \bibinfo{person}{Vaibhav Adlakha}, \bibinfo{person}{Marius Mosbach}, \bibinfo{person}{Dzmitry Bahdanau}, \bibinfo{person}{Nicolas Chapados}, {and} \bibinfo{person}{Siva Reddy}.} \bibinfo{year}{2024}\natexlab{}.
\newblock \showarticletitle{{LLM2Vec: Large Language Models Are Secretly Powerful Text Encoders}}.
\newblock \bibinfo{journal}{\emph{ArXiv}}  \bibinfo{volume}{abs/2404.05961} (\bibinfo{year}{2024}).
\newblock


\bibitem[Bonifacio et~al\mbox{.}(2022)]%
        {DBLP:journals/corr/abs-2202-05144}
\bibfield{author}{\bibinfo{person}{Luiz~Henrique Bonifacio}, \bibinfo{person}{Hugo Abonizio}, \bibinfo{person}{Marzieh Fadaee}, {and} \bibinfo{person}{Rodrigo~Frassetto Nogueira}.} \bibinfo{year}{2022}\natexlab{}.
\newblock \showarticletitle{{InPars: Data Augmentation for Information Retrieval using Large Language Models}}.
\newblock \bibinfo{journal}{\emph{ArXiv}}  \bibinfo{volume}{abs/2202.05144} (\bibinfo{year}{2022}).
\newblock


\bibitem[Craswell(2009)]%
        {DBLP:reference/db/Craswell09a}
\bibfield{author}{\bibinfo{person}{Nick Craswell}.} \bibinfo{year}{2009}\natexlab{}.
\newblock \showarticletitle{{Mean Reciprocal Rank}}.
\newblock In \bibinfo{booktitle}{\emph{Encyclopedia of Database Systems}}.
\newblock


\bibitem[Craswell et~al\mbox{.}(2021)]%
        {DBLP:conf/trec/CraswellMMYC20}
\bibfield{author}{\bibinfo{person}{Nick Craswell}, \bibinfo{person}{Bhaskar Mitra}, \bibinfo{person}{Emine Yilmaz}, {and} \bibinfo{person}{Daniel Campos}.} \bibinfo{year}{2021}\natexlab{}.
\newblock \showarticletitle{Overview of the {TREC} 2020 Deep Learning Track}. In \bibinfo{booktitle}{\emph{Text REtrieval Conference (TREC)}}.
\newblock


\bibitem[Craswell et~al\mbox{.}(2020)]%
        {DBLP:journals/corr/abs-2003-07820}
\bibfield{author}{\bibinfo{person}{Nick Craswell}, \bibinfo{person}{Bhaskar Mitra}, \bibinfo{person}{Emine Yilmaz}, \bibinfo{person}{Daniel Campos}, {and} \bibinfo{person}{Ellen~M. Voorhees}.} \bibinfo{year}{2020}\natexlab{}.
\newblock \showarticletitle{Overview of the {TREC} 2019 Deep Learning Track}. In \bibinfo{booktitle}{\emph{Text REtrieval Conference (TREC)}}.
\newblock


\bibitem[Dai et~al\mbox{.}(2023)]%
        {DBLP:conf/iclr/DaiZMLNLBGHC23}
\bibfield{author}{\bibinfo{person}{Zhuyun Dai}, \bibinfo{person}{Vincent~Y. Zhao}, \bibinfo{person}{Ji Ma}, \bibinfo{person}{Yi Luan}, \bibinfo{person}{Jianmo Ni}, \bibinfo{person}{Jing Lu}, \bibinfo{person}{Anton Bakalov}, \bibinfo{person}{Kelvin Guu}, \bibinfo{person}{Keith~B. Hall}, {and} \bibinfo{person}{Ming{-}Wei Chang}.} \bibinfo{year}{2023}\natexlab{}.
\newblock \showarticletitle{{Promptagator: Few-shot Dense Retrieval From 8 Examples}}. In \bibinfo{booktitle}{\emph{International Conference on Learning Representations (ICLR 2023)}}.
\newblock


\bibitem[Dubey et~al\mbox{.}(2024)]%
        {DBLP:journals/corr/abs-2407-21783}
\bibfield{author}{\bibinfo{person}{Abhimanyu Dubey} {et~al\mbox{.}}} \bibinfo{year}{2024}\natexlab{}.
\newblock \showarticletitle{The Llama 3 Herd of Models}.
\newblock \bibinfo{journal}{\emph{ArXiv}}  \bibinfo{volume}{abs/2407.21783} (\bibinfo{year}{2024}).
\newblock


\bibitem[Fang et~al\mbox{.}(2024)]%
        {DBLP:conf/sigir/FangZAMS0024}
\bibfield{author}{\bibinfo{person}{Yan Fang}, \bibinfo{person}{Jingtao Zhan}, \bibinfo{person}{Qingyao Ai}, \bibinfo{person}{Jiaxin Mao}, \bibinfo{person}{Weihang Su}, \bibinfo{person}{Jia Chen}, {and} \bibinfo{person}{Yiqun Liu}.} \bibinfo{year}{2024}\natexlab{}.
\newblock \showarticletitle{{Scaling Laws For Dense Retrieval}}. In \bibinfo{booktitle}{\emph{International Conference on Research and Development in Information Retrieval (SIGIR 2024)}}.
\newblock


\bibitem[Gao and Callan(2022)]%
        {DBLP:conf/acl/GaoC22}
\bibfield{author}{\bibinfo{person}{Luyu Gao} {and} \bibinfo{person}{Jamie Callan}.} \bibinfo{year}{2022}\natexlab{}.
\newblock \showarticletitle{{Unsupervised Corpus Aware Language Model Pre-training for Dense Passage Retrieval}}. In \bibinfo{booktitle}{\emph{Annual Meeting of the Association for Computational Linguistics (ACL 2022)}}.
\newblock


\bibitem[Gospodinov et~al\mbox{.}(2023)]%
        {DBLP:conf/ecir/GospodinovMM23}
\bibfield{author}{\bibinfo{person}{Mitko Gospodinov}, \bibinfo{person}{Sean MacAvaney}, {and} \bibinfo{person}{Craig Macdonald}.} \bibinfo{year}{2023}\natexlab{}.
\newblock \showarticletitle{Doc2Query-: When Less is More}. In \bibinfo{booktitle}{\emph{European Conference on Information Retrieval (ECIR 2023)}}.
\newblock


\bibitem[Han et~al\mbox{.}(2024)]%
        {han2024enhancingdenseretrieversrobustness}
\bibfield{author}{\bibinfo{person}{Peixuan Han}, \bibinfo{person}{Zhenghao Liu}, \bibinfo{person}{Zhiyuan Liu}, {and} \bibinfo{person}{Chenyan Xiong}.} \bibinfo{year}{2024}\natexlab{}.
\newblock \showarticletitle{{Enhancing Dense Retrievers' Robustness with Group-level Reweighting}}.
\newblock \bibinfo{journal}{\emph{ArXiv}}  \bibinfo{volume}{abs/2310.16605} (\bibinfo{year}{2024}).
\newblock


\bibitem[Hu et~al\mbox{.}(2022)]%
        {DBLP:conf/iclr/HuSWALWWC22}
\bibfield{author}{\bibinfo{person}{Edward~J. Hu}, \bibinfo{person}{Yelong Shen}, \bibinfo{person}{Phillip Wallis}, \bibinfo{person}{Zeyuan Allen{-}Zhu}, \bibinfo{person}{Yuanzhi Li}, \bibinfo{person}{Shean Wang}, \bibinfo{person}{Lu Wang}, {and} \bibinfo{person}{Weizhu Chen}.} \bibinfo{year}{2022}\natexlab{}.
\newblock \showarticletitle{{LoRA: Low-Rank Adaptation of Large Language Models}}. In \bibinfo{booktitle}{\emph{International Conference on Learning Representations, {(ICLR 2022)}}}.
\newblock


\bibitem[Izacard et~al\mbox{.}(2022)]%
        {DBLP:journals/tmlr/IzacardCHRBJG22}
\bibfield{author}{\bibinfo{person}{Gautier Izacard}, \bibinfo{person}{Mathilde Caron}, \bibinfo{person}{Lucas Hosseini}, \bibinfo{person}{Sebastian Riedel}, \bibinfo{person}{Piotr Bojanowski}, \bibinfo{person}{Armand Joulin}, {and} \bibinfo{person}{Edouard Grave}.} \bibinfo{year}{2022}\natexlab{}.
\newblock \showarticletitle{{Unsupervised Dense Information Retrieval with Contrastive Learning}}.
\newblock \bibinfo{journal}{\emph{Transactions on Machine Learning Research}} (\bibinfo{year}{2022}).
\newblock


\bibitem[J{\"{a}}rvelin and Kek{\"{a}}l{\"{a}}inen(2002)]%
        {DBLP:journals/tois/JarvelinK02}
\bibfield{author}{\bibinfo{person}{Kalervo J{\"{a}}rvelin} {and} \bibinfo{person}{Jaana Kek{\"{a}}l{\"{a}}inen}.} \bibinfo{year}{2002}\natexlab{}.
\newblock \showarticletitle{Cumulated gain-based evaluation of {IR} techniques}.
\newblock \bibinfo{journal}{\emph{{ACM} Transactions on Information Systems}} (\bibinfo{year}{2002}).
\newblock


\bibitem[Jeronymo et~al\mbox{.}(2023)]%
        {DBLP:journals/corr/abs-2301-01820}
\bibfield{author}{\bibinfo{person}{Vitor Jeronymo}, \bibinfo{person}{Luiz~Henrique Bonifacio}, \bibinfo{person}{Hugo Abonizio}, \bibinfo{person}{Marzieh Fadaee}, \bibinfo{person}{Roberto de Alencar~Lotufo}, \bibinfo{person}{Jakub Zavrel}, {and} \bibinfo{person}{Rodrigo~Frassetto Nogueira}.} \bibinfo{year}{2023}\natexlab{}.
\newblock \showarticletitle{{InPars-v2: Large Language Models as Efficient Dataset Generators for Information Retrieval}}.
\newblock \bibinfo{journal}{\emph{ArXiv}}  \bibinfo{volume}{abs/2301.01820} (\bibinfo{year}{2023}).
\newblock


\bibitem[Karpukhin et~al\mbox{.}(2020)]%
        {DBLP:conf/emnlp/KarpukhinOMLWEC20}
\bibfield{author}{\bibinfo{person}{Vladimir Karpukhin}, \bibinfo{person}{Barlas Oguz}, \bibinfo{person}{Sewon Min}, \bibinfo{person}{Patrick S.~H. Lewis}, \bibinfo{person}{Ledell Wu}, \bibinfo{person}{Sergey Edunov}, \bibinfo{person}{Danqi Chen}, {and} \bibinfo{person}{Wen{-}tau Yih}.} \bibinfo{year}{2020}\natexlab{}.
\newblock \showarticletitle{{Dense Passage Retrieval for Open-Domain Question Answering}}. In \bibinfo{booktitle}{\emph{Conference on Empirical Methods in Natural Language Processing (EMNLP 2020)}}.
\newblock


\bibitem[Lee et~al\mbox{.}(2024b)]%
        {DBLP:journals/corr/abs-2405-17428}
\bibfield{author}{\bibinfo{person}{Chankyu Lee}, \bibinfo{person}{Rajarshi Roy}, \bibinfo{person}{Mengyao Xu}, \bibinfo{person}{Jonathan Raiman}, \bibinfo{person}{Mohammad Shoeybi}, \bibinfo{person}{Bryan Catanzaro}, {and} \bibinfo{person}{Wei Ping}.} \bibinfo{year}{2024}\natexlab{b}.
\newblock \showarticletitle{{NV-Embed: Improved Techniques for Training LLMs as Generalist Embedding Models}}.
\newblock \bibinfo{journal}{\emph{ArXiv}}  \bibinfo{volume}{abs/2405.17428} (\bibinfo{year}{2024}).
\newblock


\bibitem[Lee et~al\mbox{.}(2024a)]%
        {DBLP:journals/corr/abs-2403-20327}
\bibfield{author}{\bibinfo{person}{Jinhyuk Lee}, \bibinfo{person}{Zhuyun Dai}, \bibinfo{person}{Xiaoqi Ren}, \bibinfo{person}{Blair Chen}, \bibinfo{person}{Daniel Cer}, \bibinfo{person}{Jeremy~R. Cole}, \bibinfo{person}{Kai Hui}, \bibinfo{person}{Michael Boratko}, \bibinfo{person}{Rajvi Kapadia}, \bibinfo{person}{Wen Ding}, \bibinfo{person}{Yi Luan}, \bibinfo{person}{Sai Meher~Karthik Duddu}, \bibinfo{person}{Gustavo~Hern{\'{a}}ndez {\'{A}}brego}, \bibinfo{person}{Weiqiang Shi}, \bibinfo{person}{Nithi Gupta}, \bibinfo{person}{Aditya Kusupati}, \bibinfo{person}{Prateek Jain}, \bibinfo{person}{Siddhartha~Reddy Jonnalagadda}, \bibinfo{person}{Ming{-}Wei Chang}, {and} \bibinfo{person}{Iftekhar Naim}.} \bibinfo{year}{2024}\natexlab{a}.
\newblock \showarticletitle{{Gecko: Versatile Text Embeddings Distilled from Large Language Models}}.
\newblock \bibinfo{journal}{\emph{ArXiv}}  \bibinfo{volume}{abs/2403.20327} (\bibinfo{year}{2024}).
\newblock


\bibitem[Lee et~al\mbox{.}(2019)]%
        {DBLP:conf/acl/LeeCT19}
\bibfield{author}{\bibinfo{person}{Kenton Lee}, \bibinfo{person}{Ming{-}Wei Chang}, {and} \bibinfo{person}{Kristina Toutanova}.} \bibinfo{year}{2019}\natexlab{}.
\newblock \showarticletitle{{Latent Retrieval for Weakly Supervised Open Domain Question Answering}}. In \bibinfo{booktitle}{\emph{Annual Meeting of the Association for Computational Linguistics (ACL 2019)}}.
\newblock


\bibitem[Lu et~al\mbox{.}(2021)]%
        {lu-etal-2021-less}
\bibfield{author}{\bibinfo{person}{Shuqi Lu}, \bibinfo{person}{Di He}, \bibinfo{person}{Chenyan Xiong}, \bibinfo{person}{Guolin Ke}, \bibinfo{person}{Waleed Malik}, \bibinfo{person}{Zhicheng Dou}, \bibinfo{person}{Paul Bennett}, \bibinfo{person}{Tie-Yan Liu}, {and} \bibinfo{person}{Arnold Overwijk}.} \bibinfo{year}{2021}\natexlab{}.
\newblock \showarticletitle{{Less is More: Pretrain a Strong {S}iamese Encoder for Dense Text Retrieval Using a Weak Decode}}. In \bibinfo{booktitle}{\emph{Conference on Empirical Methods in Natural Language Processing (EMNLP 2021)}}.
\newblock


\bibitem[Ma et~al\mbox{.}(2024)]%
        {DBLP:journals/corr/abs-2401-11248}
\bibfield{author}{\bibinfo{person}{Guangyuan Ma}, \bibinfo{person}{Xing Wu}, \bibinfo{person}{Zijia Lin}, {and} \bibinfo{person}{Songlin Hu}.} \bibinfo{year}{2024}\natexlab{}.
\newblock \showarticletitle{{Drop your Decoder: Pre-training with Bag-of-Word Prediction for Dense Passage Retrieval}}.
\newblock \bibinfo{journal}{\emph{ArXiv}}  \bibinfo{volume}{abs/2401.11248} (\bibinfo{year}{2024}).
\newblock


\bibitem[Ma et~al\mbox{.}(2022)]%
        {DBLP:conf/sigir/MaGZFC22}
\bibfield{author}{\bibinfo{person}{Xinyu Ma}, \bibinfo{person}{Jiafeng Guo}, \bibinfo{person}{Ruqing Zhang}, \bibinfo{person}{Yixing Fan}, {and} \bibinfo{person}{Xueqi Cheng}.} \bibinfo{year}{2022}\natexlab{}.
\newblock \showarticletitle{{Pre-train a Discriminative Text Encoder for Dense Retrieval via Contrastive Span Prediction}}. In \bibinfo{booktitle}{\emph{International Conference on Research and Development in Information Retrieval (SIGIR 2022)}}.
\newblock


\bibitem[Nguyen et~al\mbox{.}(2016)]%
        {DBLP:conf/nips/NguyenRSGTMD16}
\bibfield{author}{\bibinfo{person}{Tri Nguyen}, \bibinfo{person}{Mir Rosenberg}, \bibinfo{person}{Xia Song}, \bibinfo{person}{Jianfeng Gao}, \bibinfo{person}{Saurabh Tiwary}, \bibinfo{person}{Rangan Majumder}, {and} \bibinfo{person}{Li Deng}.} \bibinfo{year}{2016}\natexlab{}.
\newblock \showarticletitle{{MS} {MARCO:} {A} Human Generated MAchine Reading COmprehension Dataset}. In \bibinfo{booktitle}{\emph{Workshop on Cognitive Computation: Integrating Neural and Symbolic Approaches}}.
\newblock


\bibitem[Nogueira and Lin(2019)]%
        {nogueira2019doc2query}
\bibfield{author}{\bibinfo{person}{Rodrigo Nogueira} {and} \bibinfo{person}{Jimmy Lin}.} \bibinfo{year}{2019}\natexlab{}.
\newblock \bibinfo{booktitle}{\emph{From doc2query to docTTTTTquery}}.
\newblock \bibinfo{type}{{T}echnical {R}eport}. \bibinfo{institution}{University of Waterloo}.
\newblock


\bibitem[Nogueira et~al\mbox{.}(2019)]%
        {DBLP:journals/corr/abs-1904-08375}
\bibfield{author}{\bibinfo{person}{Rodrigo~Frassetto Nogueira}, \bibinfo{person}{Wei Yang}, \bibinfo{person}{Jimmy Lin}, {and} \bibinfo{person}{Kyunghyun Cho}.} \bibinfo{year}{2019}\natexlab{}.
\newblock \showarticletitle{{Document Expansion by Query Prediction}}.
\newblock \bibinfo{journal}{\emph{ArXiv}}  \bibinfo{volume}{abs/1904.08375} (\bibinfo{year}{2019}).
\newblock


\bibitem[Ouyang et~al\mbox{.}(2024)]%
        {DBLP:journals/corr/abs-2411-00722}
\bibfield{author}{\bibinfo{person}{Yichen Ouyang}, \bibinfo{person}{Lu Wang}, \bibinfo{person}{Fangkai Yang}, \bibinfo{person}{Pu Zhao}, \bibinfo{person}{Chenghua Huang}, \bibinfo{person}{Jianfeng Liu}, \bibinfo{person}{Bochen Pang}, \bibinfo{person}{Yaming Yang}, \bibinfo{person}{Yuefeng Zhan}, \bibinfo{person}{Hao Sun}, \bibinfo{person}{Qingwei Lin}, \bibinfo{person}{Saravan Rajmohan}, \bibinfo{person}{Weiwei Deng}, \bibinfo{person}{Dongmei Zhang}, \bibinfo{person}{Feng Sun}, {and} \bibinfo{person}{Qi Zhang}.} \bibinfo{year}{2024}\natexlab{}.
\newblock \showarticletitle{{Token-level Proximal Policy Optimization for Query Generation}}.
\newblock \bibinfo{journal}{\emph{ArXiv}}  \bibinfo{volume}{abs/2411.00722} (\bibinfo{year}{2024}).
\newblock


\bibitem[Overwijk et~al\mbox{.}(2022)]%
        {DBLP:conf/sigir/OverwijkXC22}
\bibfield{author}{\bibinfo{person}{Arnold Overwijk}, \bibinfo{person}{Chenyan Xiong}, {and} \bibinfo{person}{Jamie Callan}.} \bibinfo{year}{2022}\natexlab{}.
\newblock \showarticletitle{{ClueWeb22: 10 Billion Web Documents with Rich Information}}. In \bibinfo{booktitle}{\emph{International Conference on Research and Development in Information Retrieval (SIGIR 2022)}}.
\newblock


\bibitem[Rafailov et~al\mbox{.}(2023)]%
        {DBLP:conf/nips/RafailovSMMEF23}
\bibfield{author}{\bibinfo{person}{Rafael Rafailov}, \bibinfo{person}{Archit Sharma}, \bibinfo{person}{Eric Mitchell}, \bibinfo{person}{Christopher~D. Manning}, \bibinfo{person}{Stefano Ermon}, {and} \bibinfo{person}{Chelsea Finn}.} \bibinfo{year}{2023}\natexlab{}.
\newblock \showarticletitle{{Direct Preference Optimization: Your Language Model is Secretly a Reward Model}}. In \bibinfo{booktitle}{\emph{Conference on Neural Information Processing Systems (NeurIPS 2023)}}.
\newblock


\bibitem[Robertson and Zaragoza(2009)]%
        {DBLP:journals/ftir/RobertsonZ09}
\bibfield{author}{\bibinfo{person}{Stephen~E. Robertson} {and} \bibinfo{person}{Hugo Zaragoza}.} \bibinfo{year}{2009}\natexlab{}.
\newblock \showarticletitle{The Probabilistic Relevance Framework: {BM25} and Beyond}.
\newblock \bibinfo{journal}{\emph{Found. Trends Inf. Retr.}} (\bibinfo{year}{2009}).
\newblock


\bibitem[Schulman et~al\mbox{.}(2017)]%
        {DBLP:journals/corr/SchulmanWDRK17}
\bibfield{author}{\bibinfo{person}{John Schulman}, \bibinfo{person}{Filip Wolski}, \bibinfo{person}{Prafulla Dhariwal}, \bibinfo{person}{Alec Radford}, {and} \bibinfo{person}{Oleg Klimov}.} \bibinfo{year}{2017}\natexlab{}.
\newblock \showarticletitle{{Proximal Policy Optimization Algorithms}}.
\newblock \bibinfo{journal}{\emph{ArXiv}}  \bibinfo{volume}{abs/1707.06347} (\bibinfo{year}{2017}).
\newblock


\bibitem[Sun et~al\mbox{.}(2023)]%
        {DBLP:conf/emnlp/0001YMWRCYR23}
\bibfield{author}{\bibinfo{person}{Weiwei Sun}, \bibinfo{person}{Lingyong Yan}, \bibinfo{person}{Xinyu Ma}, \bibinfo{person}{Shuaiqiang Wang}, \bibinfo{person}{Pengjie Ren}, \bibinfo{person}{Zhumin Chen}, \bibinfo{person}{Dawei Yin}, {and} \bibinfo{person}{Zhaochun Ren}.} \bibinfo{year}{2023}\natexlab{}.
\newblock \showarticletitle{{Is ChatGPT Good at Search? Investigating Large Language Models as Re-Ranking Agents}}. In \bibinfo{booktitle}{\emph{Conference on Empirical Methods in Natural Language Processing (EMNLP 2023)}}.
\newblock


\bibitem[van~den Oord et~al\mbox{.}(2018)]%
        {DBLP:journals/corr/abs-1807-03748}
\bibfield{author}{\bibinfo{person}{A{\"{a}}ron van~den Oord}, \bibinfo{person}{Yazhe Li}, {and} \bibinfo{person}{Oriol Vinyals}.} \bibinfo{year}{2018}\natexlab{}.
\newblock \showarticletitle{{Representation Learning with Contrastive Predictive Coding}}.
\newblock \bibinfo{journal}{\emph{ArXiv}}  \bibinfo{volume}{abs/1807.03748} (\bibinfo{year}{2018}).
\newblock


\bibitem[Vaswani et~al\mbox{.}(2017)]%
        {DBLP:conf/nips/VaswaniSPUJGKP17}
\bibfield{author}{\bibinfo{person}{Ashish Vaswani}, \bibinfo{person}{Noam Shazeer}, \bibinfo{person}{Niki Parmar}, \bibinfo{person}{Jakob Uszkoreit}, \bibinfo{person}{Llion Jones}, \bibinfo{person}{Aidan~N. Gomez}, \bibinfo{person}{Lukasz Kaiser}, {and} \bibinfo{person}{Illia Polosukhin}.} \bibinfo{year}{2017}\natexlab{}.
\newblock \showarticletitle{{Attention is All you Need}}. In \bibinfo{booktitle}{\emph{Conference on Neural Information Processing Systems (NeurIPS 2017)}}.
\newblock


\bibitem[Wang et~al\mbox{.}(2022)]%
        {DBLP:journals/corr/abs-2212-03533}
\bibfield{author}{\bibinfo{person}{Liang Wang}, \bibinfo{person}{Nan Yang}, \bibinfo{person}{Xiaolong Huang}, \bibinfo{person}{Binxing Jiao}, \bibinfo{person}{Linjun Yang}, \bibinfo{person}{Daxin Jiang}, \bibinfo{person}{Rangan Majumder}, {and} \bibinfo{person}{Furu Wei}.} \bibinfo{year}{2022}\natexlab{}.
\newblock \showarticletitle{{Text Embeddings by Weakly-Supervised Contrastive Pre-training}}.
\newblock \bibinfo{journal}{\emph{ArXiv}}  \bibinfo{volume}{abs/2212.03533} (\bibinfo{year}{2022}).
\newblock


\bibitem[Wang et~al\mbox{.}(2024)]%
        {DBLP:conf/acl/WangYHYMW24}
\bibfield{author}{\bibinfo{person}{Liang Wang}, \bibinfo{person}{Nan Yang}, \bibinfo{person}{Xiaolong Huang}, \bibinfo{person}{Linjun Yang}, \bibinfo{person}{Rangan Majumder}, {and} \bibinfo{person}{Furu Wei}.} \bibinfo{year}{2024}\natexlab{}.
\newblock \showarticletitle{{Improving Text Embeddings with Large Language Models}}. In \bibinfo{booktitle}{\emph{Annual Meeting of the Association for Computational Linguistics (ACL 2024)}}.
\newblock


\bibitem[Xiao et~al\mbox{.}(2022)]%
        {DBLP:conf/emnlp/XiaoLSC22}
\bibfield{author}{\bibinfo{person}{Shitao Xiao}, \bibinfo{person}{Zheng Liu}, \bibinfo{person}{Yingxia Shao}, {and} \bibinfo{person}{Zhao Cao}.} \bibinfo{year}{2022}\natexlab{}.
\newblock \showarticletitle{{RetroMAE: Pre-Training Retrieval-oriented Language Models Via Masked Auto-Encoder}}. In \bibinfo{booktitle}{\emph{Conference on Empirical Methods in Natural Language Processing (EMNLP 2022)}}.
\newblock


\bibitem[Xiong et~al\mbox{.}(2020)]%
        {DBLP:journals/corr/abs-2011-01580}
\bibfield{author}{\bibinfo{person}{Chenyan Xiong}, \bibinfo{person}{Zhenghao Liu}, \bibinfo{person}{Si Sun}, \bibinfo{person}{Zhuyun Dai}, \bibinfo{person}{Kaitao Zhang}, \bibinfo{person}{Shi Yu}, \bibinfo{person}{Zhiyuan Liu}, \bibinfo{person}{Hoifung Poon}, \bibinfo{person}{Jianfeng Gao}, {and} \bibinfo{person}{Paul Bennett}.} \bibinfo{year}{2020}\natexlab{}.
\newblock \showarticletitle{{CMT} in {TREC-COVID} Round 2: Mitigating the Generalization Gaps from Web to Special Domain Search}.
\newblock \bibinfo{journal}{\emph{ArXiv}}  \bibinfo{volume}{abs/2011.01580} (\bibinfo{year}{2020}).
\newblock


\bibitem[Xiong et~al\mbox{.}(2021)]%
        {DBLP:conf/iclr/XiongXLTLBAO21}
\bibfield{author}{\bibinfo{person}{Lee Xiong}, \bibinfo{person}{Chenyan Xiong}, \bibinfo{person}{Ye Li}, \bibinfo{person}{Kwok{-}Fung Tang}, \bibinfo{person}{Jialin Liu}, \bibinfo{person}{Paul~N. Bennett}, \bibinfo{person}{Junaid Ahmed}, {and} \bibinfo{person}{Arnold Overwijk}.} \bibinfo{year}{2021}\natexlab{}.
\newblock \showarticletitle{{Approximate Nearest Neighbor Negative Contrastive Learning for Dense Text Retrieval}}. In \bibinfo{booktitle}{\emph{International Conference on Learning Representations (ICLR 2021)}}.
\newblock


\bibitem[Yang et~al\mbox{.}(2024)]%
        {DBLP:journals/corr/abs-2412-15115}
\bibfield{author}{\bibinfo{person}{An Yang} {et~al\mbox{.}}} \bibinfo{year}{2024}\natexlab{}.
\newblock \showarticletitle{{Qwen2.5 Technical Report}}.
\newblock \bibinfo{journal}{\emph{ArXiv}}  \bibinfo{volume}{abs/2412.15115} (\bibinfo{year}{2024}).
\newblock


\end{thebibliography}

% \appendix





\end{document}